\documentclass[sigconf,natbib=true]{acmart}
\AtBeginDocument{%
  }

\copyrightyear{2026}
\acmYear{2026}
\setcopyright{cc}
\setcctype{by}
\acmConference[SIGIR '26]{Proceedings of the 49th International ACM SIGIR Conference on Research and Development in Information Retrieval}{July 20--24, 2026}{Melbourne, VIC, Australia}
\acmBooktitle{Proceedings of the 49th International ACM SIGIR Conference on Research and Development in Information Retrieval (SIGIR '26), July 20--24, 2026, Melbourne, VIC, Australia}
\acmDOI{10.1145/3805712.3809939}
\acmISBN{979-8-4007-2599-9/2026/07}

\usepackage{caption}
\usepackage{geometry}
\usepackage{multirow}
\usepackage{colortbl}
\usepackage{xcolor} 

\usepackage{subcaption}
\usepackage{lipsum}
\usepackage[font=small]{caption}
\begin{document}

\title{Improving Ad-hoc Search Effectiveness for Conversational Information Retrieval via Model Merging}

\author{Ahmed Rayane Kebir}
\orcid{0009-0009-2512-832X}
\affiliation{%
  \institution{University of Toulouse, IRIT}
  \city{Toulouse}
  \country{France}
}
\email{ahmed-rayane.kebir@irit.fr}

\author{Jose G. Moreno}
\orcid{0000-0002-8852-5797}
\affiliation{%
  \institution{University of Toulouse, IRIT}
  \city{Toulouse}
  \country{France}}
\email{jose.moreno@irit.fr}

\author{Lynda Tamine}
\orcid{0000-0002-3615-8032}
\affiliation{%
  \institution{University of Toulouse, IRIT}
  \city{Toulouse}
  \country{France}}
  \email{lynda.tamine@irit.fr}


\begin{abstract}

Conversational information retrieval is challenging since it requires the consideration of the conversation history which potentially gives rise to topic shifts and coreference resolution across previous turns. To address these challenges, previous work mainly rely on traditional fine-tuning of ad-hoc retrievers on conversational datasets or extrapolates their generalizability through multi-tasking. However, this mainstream approach is costly—since it requires model re-training—and exhibits catastrophic forgetting, where the model loses its foundational ad-hoc retrieval performance. In this paper, we fill this gap by introducing model merging as a training-free strategy enabling the design of a single retrieval model that operates across both ad-hoc and conversational settings with no additional fine-tuning. 
 We conduct experiments using linear and non-linear parameter-wise merging strategies—namely Model Soup and Slerp—on standard ad-hoc search and conversational retrieval datasets. Our results demonstrate that model merging significantly enhances the ad-hoc search capabilities of conversational retrievers while improving generalizability across task-specific datasets, achieving up to 15\% higher NDCG@3 under zero-shot conditions.

\end{abstract}



\maketitle

\section{Introduction}

Conversational search is a well-established paradigm in which users engage in natural language, multi-turn interactions with a system to satisfy evolving information needs \cite{Radlinski17,mo2025survey}. A core component of conversational search is retrieving relevant documents by leveraging the conversational context in response to the user’s latest utterance. Compared to traditional ad-hoc retrieval, conversational information retrieval (CIR) introduces additional challenges. As interactions progress, the conversational history becomes longer and noisier, often containing topic shifts, coreference, and linguistic ambiguities. Consequently, the latest user turn may be underspecified or context-dependent, increasing the difficulty of interpreting the information need and retrieving relevant documents \cite{gao2023neural}.

\begin{figure}[th]
    \centering
    \includegraphics[height=0.3\linewidth]{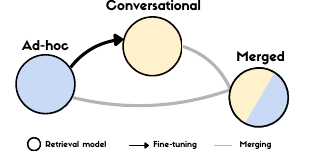}
    \caption{Model merging of a base ad-hoc retriever and a conversational fine-tuned model following   a merging pattern presented by Matena et al.\cite{matena2022merging}.}
    \label{fig:methodology}
\end{figure}

To tackle these challenges, prior work has explored a variety of approaches for training dense retrievers tailored to conversational retrieval. Common methods typically fine-tune retrievers to operate over conversational sessions that include the current query, previous turns, system responses, and additional contextual signals \cite{mo2024aligning,mao2023learning,mao2024chatretriever}. 
Given the scarcity of large-scale conversational IR datasets, most approaches adapt pretrained ad-hoc retrievers—originally trained on large datasets such as MS MARCO— to conversational contexts \cite{mo2024aligning,yu2021few,lin-etal-2021-contextualized}. 
However, task-specific adaptation for conversational retrieval often degrades performance on standard ad-hoc retrieval tasks. Mo et al. \cite{mo2025survey} report that retrievers fine-tuned on conversational datasets lose their general ad-hoc retrieval capabilities,
and emphasize the need for retrievers that generalize across ad-hoc and conversational settings. Similar observations of catastrophic forgetting in ad-hoc retrieval performance are also reported by Yang et al. \cite{yang-etal-2025-learning}.
A common solution is multi-task learning \cite{mao2024chatretriever}, where retrievers are jointly trained on ad-hoc and conversational datasets or optimized with multiple task-specific losses. However, such approaches require retraining on large-scale ad-hoc datasets, which is computationally expensive. \\
In this work, we investigate model merging as a training-free alternative to multi-task fine-tuning for CIR. Unlike multi-task learning, which jointly updates model parameters using task-specific data, model merging aggregates the parameters of independently fine-tuned models to transfer complementary capabilities without additional optimization \cite{zhou-etal-2025-mergeme,sun-etal-2025-personality, sasaki2025effect, braga2025investigating}. Our motivation is inspired by robust fine-tuning, originally introduced in the computer vision literature by Wortsman et al. \cite{wortsman2022model}, to address the trade-off between in-domain accuracy and robustness to distribution shift. In that setting, interpolating the weights of a pre-trained model and its fine-tuned counterpart mitigates over-specialization caused by fine-tuning on a narrow target distribution. We observe a similar phenomenon in conversational retrieval: fine-tuning dense retrievers on long, noisy conversational contexts induces specialization that degrades foundational ad-hoc retrieval performance. As illustrated in Fig. ~\ref{fig:methodology}, rather than applying additional or constrained fine-tuning, we adopt a training-free model merging strategy that interpolates a backbone ad-hoc retriever with its conversational fine-tuned version, unifying robustness and conversational task-specific specialization within a single model.

In this paper, we answer  the following research questions: 
\\
\textbf{RQ1:} Can model merging construct conversational retrievers with improved ad-hoc retrieval capabilities?\\
\textbf{RQ2:} Is model merging a comparable training-free alternative to multi-task fine-tuning and early stopping strategies?\\
\textbf{RQ3:} To what extent does model merging lead to conversational retrievers generalizable across datasets and tasks?

To answer these research questions, we explore merging a state-of-the-art conversational model, QRACDR \cite{mo2024aligning}, with its base ad-hoc search model, ANCE \cite{xiongapproximate}. We investigate the parameter space created by up to $80$ merge configurations and evaluate the resulting retrievers on both ad-hoc (MS MARCO) and conversational (QReCC, TopiOCQA) datasets. A particular emphasis is placed on out-of-domain (OOD) generalizability to the CAsT, NQ, and HotpotQA datasets. Our findings demonstrate that model merging achieves results comparable to multi-task fine-tuning  while effectively mitigating catastrophic forgetting. Notably, model merging eliminates the 8.47\% loss observed when using rewrite input and achieves up to 10.11\% higher NDCG@3 in session-based retrieval on CAsT.


\begin{table*}[h]
\centering
\caption{OOD retrieval performance on conversational and ad-hoc datasets. Relative changes are reported for conversational improvement ($\delta_i$) with respect to the base conversational model and ad-hoc forgetting ($\delta_f$) with respect to the ANCE baseline (NDCG@3), ${^{*}, ^{\boldsymbol{\dagger}}, ^{\boldsymbol{\ddagger}}}$ marks statistically significant improvements over ANCE, QRACDR-T and QRACDR-Q respectively, determined by a paired t-test with p<0.05.}

\label{tab:retriever_results_ndcg_only}
\resizebox{\linewidth}{!}{%
\setlength{\tabcolsep}{2pt}
\begin{tabular}{ll cccc cccc c cccc cccc}
\toprule
\multicolumn{2}{c}{} & \multicolumn{8}{c}{\textbf{Conversational}} & & \multicolumn{8}{c}{\textbf{Ad-hoc}} \\
\cmidrule(lr){3-10} \cmidrule(lr){12-19}
& & \multicolumn{4}{c}{\textbf{CAsT-19}} & \multicolumn{4}{c}{\textbf{CAsT-20}} & & \multicolumn{4}{c}{\textbf{HotpotQA}} & \multicolumn{4}{c}{\textbf{NQ}} \\
\cmidrule(lr){3-6} \cmidrule(lr){7-10} \cmidrule(lr){12-15} \cmidrule(lr){16-19}
& \textbf{Method} & \small MRR & \small NDCG@3 & \small R@10 & \small $\delta_i$ & \small MRR & \small NDCG@3 & \small R@10 & \small $\delta_i$ & & \small MRR & \small NDCG@3 & \small R@10 & \small $\delta_f$ & \small MRR & \small NDCG@3 & \small R@10 & \small $\delta_f$ \\
\midrule
\multicolumn{19}{c}{\textbf{Base Ad-hoc Models}} \\
\midrule
& ANCE \cite{xiongapproximate} & 42.32 & 26.28 & 9.80 & -- & 29.04 & 18.60 & 7.23 & -- & & 65.91 & 43.48 & 47.15 & -- & 34.42 & 31.00 & 52.56 & -- \\
\midrule
\multicolumn{19}{c}{\textbf{Base Conversational Models}} \\
\midrule
& QRACDR-T \cite{mo2024aligning}& 59.46 & 41.02 & 14.92 & -- & 40.36 & 26.69 & 13.02 & -- & & 62.81 & 41.00 & 44.56 & -5.70 & 34.46 & 30.88 & 51.01 & -0.39 \\
& QRACDR-Q \cite{mo2024aligning}& 55.54 & 37.88 & 13.07 & -- & 37.32 & 22.65 & 11.01 & -- & & 51.46 & 32.83 & 38.52 & -24.49 & 28.43 & 24.85 & 45.22 & -19.84 \\
\midrule
\multicolumn{19}{c}{\textbf{Merged Models (ours)}} \\
\midrule
& QRACDR-T$_{MG-SL}$ & 58.88$^{\boldsymbol{*}}$ & 41.52$^{\boldsymbol{*}}$ & 13.87$^{\boldsymbol{*}}$ & \textbf{+1.22} & 42.45$^{\boldsymbol{*}\boldsymbol{\ddagger}}$ & 27.83$^{{\boldsymbol{*}}\boldsymbol{\ddagger}}$ & 12.29$^{\boldsymbol{*}}$ & \textbf{+4.27} & & 66.74$^{\boldsymbol{*}\boldsymbol{\dagger}\boldsymbol{\ddagger}}$ & 44.02$^{\boldsymbol{*}\boldsymbol{\dagger}\boldsymbol{\ddagger}}$ & 47.77$^{\boldsymbol{*}\boldsymbol{\dagger}\boldsymbol{\ddagger}}$ & \textbf{+1.24} & 35.64$^{\boldsymbol{*}\boldsymbol{\dagger}\boldsymbol{\ddagger}}$ & 32.19$^{\boldsymbol{*}\boldsymbol{\dagger}\boldsymbol{\ddagger}}$ & 53.12$^{\boldsymbol{\dagger}\boldsymbol{\ddagger}}$ & \textbf{+3.84} \\
& QRACDR-Q$_{MG-SL}$    & 55.22$^{\boldsymbol{*}}$ & 36.65$^{\boldsymbol{*}}$ & 12.90$^{\boldsymbol{*}}$ & -3.24 & 37.98$^{\boldsymbol{*}}$ & 24.58$^{\boldsymbol{*}}$ & 11.09$^{\boldsymbol{*}}$ & +8.52 & & 62.37$^{\boldsymbol{\ddagger}}$ & 40.95$^{\boldsymbol{\ddagger}}$ & 45.61$^{\boldsymbol{\dagger}\boldsymbol{\ddagger}}$ & \textbf{-5.82} & 33.25$^{\boldsymbol{\ddagger}}$ & 29.63$^{\boldsymbol{\ddagger}}$ & 51.42$^{\boldsymbol{\ddagger}}$ & \textbf{-4.42} \\
\midrule
& QRACDR-T$_{\text{MG}-MS}$ & 58.63$^{\boldsymbol{*}}$ & 39.90$^{\boldsymbol{*}}$ & 14.42$^{\boldsymbol{*}\boldsymbol{\ddagger}}$ & -2.73 & 41.77$^{\boldsymbol{*}\boldsymbol{\ddagger}}$ & 27.79$^{\boldsymbol{*}\boldsymbol{\ddagger}}$ & 11.79$^{\boldsymbol{*}}$ & +4.12 & & 66.36$^{\boldsymbol{\dagger}\boldsymbol{\ddagger}}$ & 43.72$^{\boldsymbol{\dagger}\boldsymbol{\ddagger}}$ & 47.79$^{\boldsymbol{*}\boldsymbol{\dagger}\boldsymbol{\ddagger}}$ & +0.55 & 35.90$^{\boldsymbol{*}\boldsymbol{\dagger}\boldsymbol{\ddagger}}$ & 32.42$^{\boldsymbol{*}\boldsymbol{\dagger}\boldsymbol{\ddagger}}$ & 53.44$^{\boldsymbol{*}\boldsymbol{\dagger}\boldsymbol{\ddagger}}$ & +4.58 \\
& QRACDR-Q$_{\text{MG}-MS}$ & 57.11$^{\boldsymbol{*}}$ & 37.80$^{\boldsymbol{*}}$ & 13.59$^{\boldsymbol{*}}$ & -0.21 & 37.79$^{\boldsymbol{*}}$ & 24.94$^{\boldsymbol{*}\boldsymbol{\dagger}}$ & 11.05$^{\boldsymbol{*}}$ & \textbf{+10.11} & & 62.01$^{\boldsymbol{\ddagger}}$ & 40.49$^{\boldsymbol{\ddagger}}$ & 45.11$^{\boldsymbol{\dagger}\boldsymbol{\ddagger}}$ & -6.88 & 33.05$^{\boldsymbol{\ddagger}}$ & 29.52$^{\boldsymbol{\ddagger}}$ & 51.04$^{\boldsymbol{\ddagger}}$ & -4.77 \\
\bottomrule
\end{tabular}}
\end{table*}

\section{Related Work}

\paragraph{\textbf{Conversational Search}}
Research in conversational search primarily follows two directions: \emph{conversational query rewriting} (CQR) \cite{mo2024chiq,mao2023large}, which de-contextualizes queries for use with ad-hoc retrievers, and \emph{conversational dense retrieval} (CDR) \cite{mo2024history,mao2024chatretriever,yang-etal-2025-learning}, which performs end-to-end retrieval by encoding the full session context into dense representations.
Most CDR approaches build upon well-trained ad-hoc dense retrievers and adapt them to conversational settings through fine-tuning on conversational search datasets. Typical techniques include contrastive learning with positive and negative passages  \cite{yu2021few}, where conversational context is incorporated into the query encoder to better model session-level intent.
Several studies further improve CDR by addressing conversational noise through history selection or context denoising, selectively incorporating only the most informative turns from the conversation history \cite{yu2021few,mo2024history}. Other work explores the integration of signals from both CQR and CDR, aligning rewritten queries with dense conversational representations to improve retrieval robustness \cite{mo2024aligning}.
\paragraph{\textbf{Model Merging}}

Model merging is a training-free approach for combining task-specific knowledge, motivated by the observation that fine-tuned models for related tasks often occupy compatible regions of the parameter space \cite{matena2022merging}. A common strategy is to linearly interpolate model parameters; Wortsman et al.~\cite{wortsman2022model} demonstrated its effectiveness by merging multiple models in image classification tasks. Beyond linear interpolation, Slerp~\cite{jang2024spherical} performs interpolation along a spherical path to better preserve model norms.
In Information Retrieval (IR), prior work explored task arithmetic ~\cite{ilharcoediting} and linear interpolation for domain adaptation, merging a general-purpose IR model with domain-specific models to improve retrieval performance in the target domain \cite{sasaki2025effect, braga2025investigating}. 
To the best of our knowledge, our work is the first attempt to explore model merging to fill a research gap in conversational search.

\section{The Merging Problem Definition}

Let us consider a backbone parametric model $\mathcal{M}_{\theta^0}$ used for our tasks of interest, ad-hoc and conversational retrieval, defined below:

\paragraph{\textbf{Ad-hoc search task.}} Given a standalone query $q$, the goal of this task is to retrieve a ranked list of documents
$D = \{d_1, d_2, \ldots, d_n\}$ from a document collection $\mathcal{C}$.
This task is inherently single-turn and context-free. We consider $\mathcal{M}_{\theta^{adh}}$ as the model fine-tuned for ad-hoc search using a training dataset $\mathcal{D}^{adh}$.

\paragraph{\textbf{\textit{CIR task}}.} The goal of the CIR task is to fulfill the user's information at each turn of interaction by retrieving for the current turn $k$, the ranked list of documents $D_k = \{d_{k1}, d_{k2}, \ldots, d_{kn}\}$  from a collection $\mathcal{C}$, that better answer query at the current turn $q_k$ and history context $H_{k-1}  \ = \{(q_i, \ r_i, \ D_i )\}_{i=0}^{k-1}$ of the previous turns $1..k-1$, where $q_i$ is the user query at turn $i<k$, $r_i$ is the generated answer at turn $i$, and $D_i$ is the list of retrieved documents for the same turn. We consider $\mathcal{M}_{\theta^{cir}}$ as the model fine-tuned for ad-hoc search using a training dataset $\mathcal{D}^{cir}$.

\paragraph{\textbf{\textit{Problem}}}
We aim to build a single effective model performing both tasks, ad-hoc search and CIR. We propose a solution based on model merging \cite{jang2024spherical,wortsman2022model}.

Formally, model merging seeks to compute a merged parameter set $\theta^{mg}$
by combining the parameters of the ad-hoc and conversational models using a merging function $f$:
\begin{equation}
\theta^{mg} = f(\boldsymbol{\lambda}, \theta^{adh}, \theta^{cir}),
\end{equation}
where $\boldsymbol{\lambda}$ is a vector of interpolation coefficients applied depth-wise across the encoder layers. Each element of $\boldsymbol{\lambda}$ controls the relative contribution of the two source models at a given depth, enabling non-uniform interpolation throughout the network. We explore the following core merging functions $f$:

\noindent -\textit{Model Soup (MS)} \cite{wortsman2022model}: Consists of performing a linear combination of input models' weights using a model-wise coefficient. Formally $\theta^{mg}=\boldsymbol{\lambda}\theta^{cir} + (1-\boldsymbol{\lambda})\theta^{adh}$.

\noindent -\textit{Slerp (SL)} \cite{jang2024spherical}: Spherical Linear Interpolation is based on the angular combination of the input models such that:
$
\theta^{mg} =
\frac{\sin\!\big(\boldsymbol{\lambda}\,\Omega\big)}{\sin(\Omega)} \,\theta^{cir} + \frac{\sin\!\big((1-\boldsymbol{\lambda})\,\Omega\big)}{\sin(\Omega)} \,\theta^{adh},
$
where $\Omega$ denotes the angle between the parameter vectors $\theta^{cir}$ and $\theta^{adh}$, defined as
$
\cos(\Omega) = \frac{\langle \theta^{cir}, \theta^{adh} \rangle}{\|\theta^{cir}\| \, \|\theta^{adh}\|}.
$
The interpolation is applied depth-wise across encoder layers, with each element of $\boldsymbol{\lambda}$ controlling the contribution of the two source models at the corresponding layer.

\section{Experiments and Results}
\subsection{Experimental Setup}
\paragraph{Datasets and Metrics}

We evaluate our approach on seven publicly available datasets. These include three ad-hoc retrieval datasets from the BEIR benchmark \cite{thakurbeir}: MS MARCO \cite{nguyen2016ms}, which is considered in-domain since the base ad-hoc model is trained on its training split, NQ \cite{kwiatkowski2019natural}, and HotpotQA \cite{yang2018hotpotqa}, which are treated as OOD datasets to assess generalization. In addition, we use four standard conversational search datasets: QReCC \cite{anantha2021open} and TopiOCQA \cite{adlakha2022topiocqa} are used to train conversational retrievers and therefore serve as in-domain evaluation for conversational retrieval. CAsT 2019 \cite{dalton2020cast} and CAsT 2020 \cite{dalton2021cast} datasets are employed for OOD generalizability. Following established evaluation protocols in ad-hoc and conversational search \cite{xiongapproximate, mo2024aligning}, we measure retrieval effectiveness using three metrics: MRR, NDCG, and R@10. We specifically report NDCG@3 to emphasize top-ranked precision, which is particularly important in conversational search scenarios. All metrics are computed using the \texttt{pytrec\_eval} \cite{van2018pytrec_eval}.
Significance tests are conducted using paired t-tests at a p-value $p< 0.05$ level.

\paragraph{Models}
We employ ANCE \cite{xiongapproximate}, an encoder-only bi-encoder pre-trained on the MS MARCO ad-hoc retrieval task, as our base dense retriever (i.e., $\mathcal{M}_{\theta^{adh}}$). For conversational retrieval, we use QRACDR \cite{mo2024aligning}, which results from fine-tuning ANCE models specifically for conversational contexts (i.e., $\mathcal{M}_{\theta^{cir}}$). This workflow follows the merging pattern presented in Fig. ~\ref{fig:methodology}. Since the original weights are not public, we used the authors' shared code\footnote{We followed the authors' implementation \url{https://github.com/fengranMark/QRACDR}.} to reproduce the fine-tuned versions of the models, noted QRACDR-Q, QRACDR-T, on the QReCC and TopiOCQA datasets, respectively. To create our merged models, we combine the base ANCE weights with those of the fine-tuned QRACDR variants using the MergeKit \cite{goddard2024arcee}. We denote the resulting models as $\text{QRACDR-Q}_{MG}$ and $\text{QRACDR-T}_{MG}$. Specifically, we distinguish between the merging strategies as $\text{QRACDR-Q}_{MG-MS}$ and QRACDR-T$_{MG-MS}$ for Model Soup (MS), and $\text{QRACDR-Q}_{MG\text{-}SL}$ and $\text{QRACDR-T}_{MG\text{-}SL}$ for Slerp model (SL). The code used to generate these merges is available at \url{https://github.com/RayaneA7/model-merging-CIR}.

\paragraph{Merging Optimization}
The optimization of the interpolation vector $\boldsymbol{\lambda}$ (§ Eq. 1) is conducted exclusively on \emph{in-domain} datasets corresponding to each task: MS MARCO for ad-hoc retrieval, and QReCC or TopiOCQA for conversational retrieval. All OOD datasets are strictly reserved for evaluation and are not used during coefficient selection. We generated 40 configurations using MS and 40 using SL merging models, and we then  selected a final merged model for each merging strategy. The selected configurations employ depth-wise interpolation vectors $\boldsymbol{\lambda}^\star$  such as $\boldsymbol{\lambda}^\star_{MS} = (1.0, 0.73, 0.47, 0.2)$ for MS and $\boldsymbol{\lambda}^\star_{SL} = (0.5, 0.6, 0.6, 0.5)$ using SL, applied from lower to higher encoder layers.

\subsection{Results}

\subsubsection{\textbf{Can model merging construct conversational retrievers with improved ad-hoc retrieval capabilities? (RQ1)}}

\begin{figure}[htbp]
    \centering
    \begin{subfigure}[b]{0.45\textwidth}
        \centering
        \includegraphics[width=\linewidth]{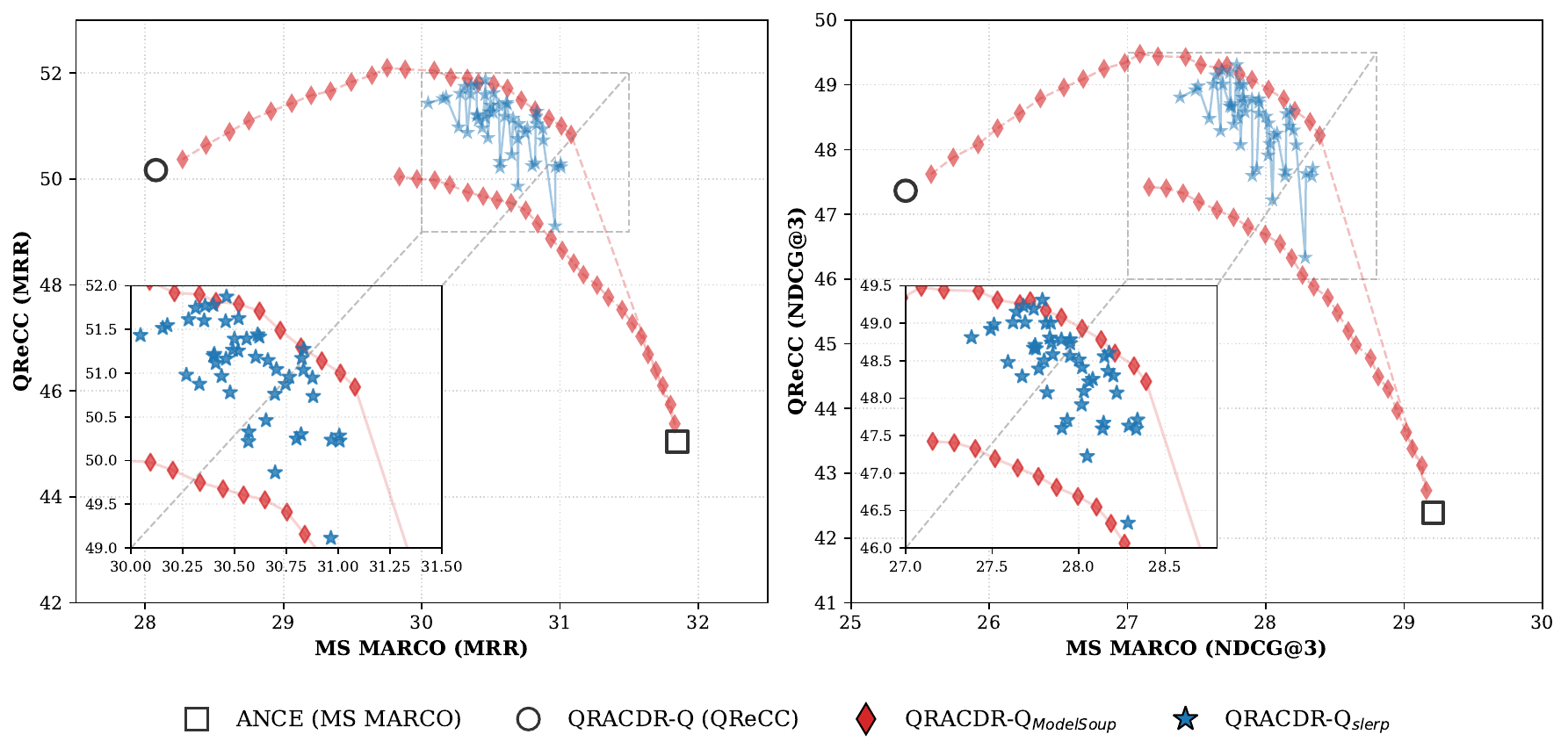}
        \caption{QReCC}
    \end{subfigure}
\hfill
   \begin{subfigure}[b]{0.45\textwidth}
        \centering
        \includegraphics[width=\linewidth]{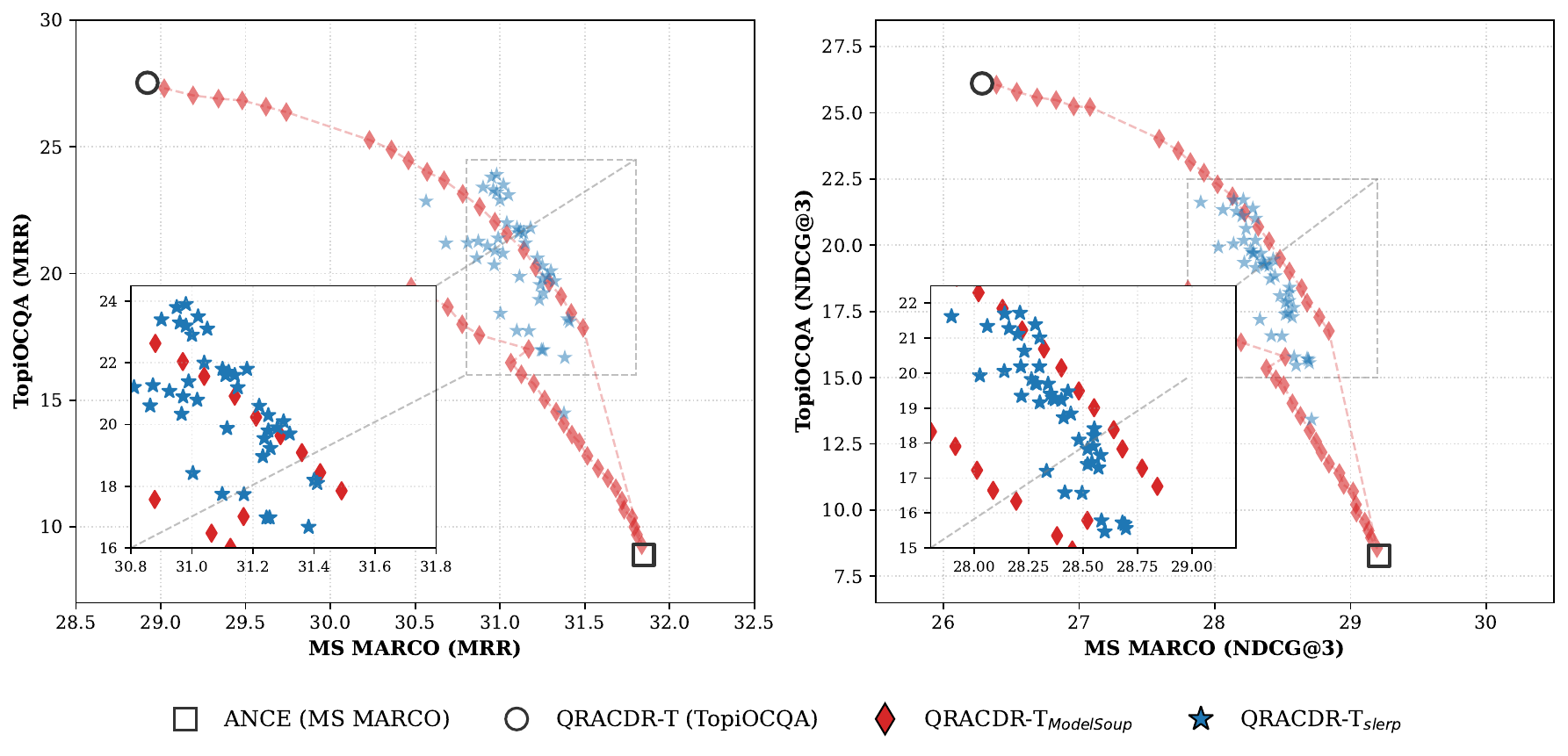}  
        \caption{TopiOCQA}
    \end{subfigure}    
    \caption{Performance of a total of 80 merged models (40 for MS and 40 for SL) by varying parameter $\lambda$, compared to base retriever models (ANCE) and conversational retriever (QRACDR) fine-tuned on QRECC (top), and TopiocQA (bottom). Results are reported using MRR (left) and NDCG (right) metrics. The optimal performances of ANCE on the MSMARCO and QRACDR on QRECC and TopiocQA  are noted respectively with square and circle symbols. }
    \label{fig:merging-space}
\end{figure}

Fig. ~\ref{fig:merging-space} shows that the conversational models QRACDR-Q and QRACDR-T achieve, as expected, the highest performance on their respective in-domain conversational test sets QReCC and TopiOCQA. However, this improvement comes at the cost of degraded ad-hoc retrieval performance on MS MARCO when compared to the base \texttt{ANCE} model (-14\% in NDCG@3 for QRACDR-Q). This observation confirms that conversational fine-tuning induces a catastrophic forgetting effect on ad-hoc retrieval capabilities.

We interestingly observe that both linear MS and non-linear SL merging strategies produce models whose performance lies between that of the base ad-hoc retriever and the conversational fine-tuned retrievers. Models in this merged space consistently recover a large portion of the ad-hoc effectiveness on MS MARCO, while incurring only a moderate degradation in conversational retrieval performance, for instance QRACDR-Q$_{MG-MS}$ reduced NDCG@3 decrease on MS MARCO from -14\% to -3\% while keeping similar performance on QReCC as QRACDR-Q.

These results indicate that model merging can effectively restore ad-hoc capabilities in conversational retrievers with no further training cost. In addition, we observe that a large subset of QRACDR-Q$_{MG}$ models improves performance on the conversational dataset QReCC itself (4\% increase in NDCG@3 w.r.t QRACDR-Q), suggesting that preserving ad-hoc retrieval capabilities can also benefit conversational retrieval. In contrast, this trend is not observed for ANCE models merged with QRACDR-T. We attribute this difference to intrinsic properties of TopiOCQA, which contains more frequent topic shifts and longer, more complex conversational queries than QReCC. These characteristics encourage stronger specialization toward conversational context, limiting the positive transfer from ad-hoc representations.

\begin{table*}[!ht]
\caption{OOD performance on CAsT using Session or manual rewritten query as input. $\delta_i$ denotes Session improvement vs. base conversational; $\delta_f$ denotes Rewrite forgetting vs. ANCE base (NDCG@3), ${^{*}, ^{\boldsymbol{\dagger}}, ^{\boldsymbol{\ddagger}}}$ marks statistically significant improvements over ANCE, QRACDR-T and QRACDR-Q respectively, determined by a paired t-test with p<0.05.}

\label{tab:rewrite_perf}
\centering
\resizebox{\linewidth}{!}{%
\setlength{\tabcolsep}{2pt}
\begin{tabular}{ll cccc cccc c cccc cccc}
\toprule
\multicolumn{2}{c}{} & \multicolumn{8}{c}{\textbf{CAsT-19}} & & \multicolumn{8}{c}{\textbf{CAsT-20}} \\
\cmidrule(lr){3-10} \cmidrule(lr){12-19}
& & \multicolumn{4}{c}{\textbf{Session}} & \multicolumn{4}{c}{\textbf{Rewrite}} & & \multicolumn{4}{c}{\textbf{Session}} & \multicolumn{4}{c}{\textbf{Rewrite}} \\
\cmidrule(lr){3-6} \cmidrule(lr){7-10} \cmidrule(lr){12-15} \cmidrule(lr){16-19}
& \textbf{Method} & \small MRR & \small NDCG@3 & \small R@10 & \small $\delta_i$ & \small MRR & \small NDCG@3 & \small R@10 & \small $\delta_f$ & & \small MRR & \small NDCG@3 & \small R@10 & \small $\delta_i$ & \small MRR & \small NDCG@3 & \small R@10 & \small \textbf{$\delta_f$} \\
\midrule
\multicolumn{19}{c}{\textbf{Base Ad-hoc Models}} \\
\midrule
& ANCE \cite{xiongapproximate} & 42.32 & 26.28 & 9.80 & -- & 66.30 & 46.89 & 16.30 & -- & & 29.04 & 18.60 & 7.23 & -- & 60.62 & 44.47 & 20.77 & -- \\
\midrule
\multicolumn{19}{c}{\textbf{Base Conversational Models }} \\
\midrule
& QRACDR-T \cite{mo2024aligning} & 59.46 & 41.02 & 14.92 & -- & 60.72 & 42.92 & 14.70 & -8.47 & & 40.36 & 26.69 & 13.02 & -- & 57.90 & 39.95 & 19.74 & -10.16 \\
& QRACDR-Q \cite{mo2024aligning} & 55.54 & 37.88 & 13.07 & -- & 56.96 & 39.44 & 13.94 & -15.89 & & 37.32 & 22.65 & 11.01 & -- & 53.48 & 39.01 & 18.86 & -12.28 \\
\midrule
\multicolumn{19}{c}{\textbf{Merged Models (ours)}} \\
\midrule
& QRACDR-T$_{MG-SL}$ & 58.88$^{\boldsymbol{*}}$ & 41.52$^{\boldsymbol{*}}$ & 13.87$^{\boldsymbol{*}}$ & \textbf{+1.22} & 64.99$^{\boldsymbol{\dagger}\boldsymbol{\ddagger}}$ & 45.81$^{\boldsymbol{\dagger}\boldsymbol{\ddagger}}$ & 15.94$^{\boldsymbol{\dagger}\boldsymbol{\ddagger}}$ & -2.30 & & 42.45$^{\boldsymbol{*}\boldsymbol{\ddagger}}$ & 27.83$^{\boldsymbol{*}\boldsymbol{\ddagger}}$ & 12.29$^{\boldsymbol{*}}$ & \textbf{+4.27} & 61.99$^{\boldsymbol{\dagger}\boldsymbol{\ddagger}}$ & 44.17$^{\boldsymbol{\dagger}\boldsymbol{\ddagger}}$ & 21.25$^{\boldsymbol{\dagger}\boldsymbol{\ddagger}}$ & -0.67 \\
& QRACDR-Q$_{MG-SL}$    & 55.22$^{\boldsymbol{*}}$ & 36.65$^{\boldsymbol{*}}$ & 12.90$^{\boldsymbol{*}}$ & -3.24 & 60.91$^{\boldsymbol{\ddagger}}$ & 43.88$^{\boldsymbol{\ddagger}}$ & 15.41$^{\boldsymbol{\ddagger}}$ & -6.42 & & 37.98$^{\boldsymbol{*}}$ & 24.58$^{\boldsymbol{*}}$ & 11.09$^{\boldsymbol{*}}$ & +8.52 & 59.11$^{\boldsymbol{\ddagger}}$ & 43.13$^{\boldsymbol{\dagger}\boldsymbol{\ddagger}}$ & 20.44$^{\boldsymbol{\ddagger}}$ & -3.01 \\
\midrule
&  QRACDR-T$_{MG-MS}$ & 58.63$^{\boldsymbol{*}}$ & 39.90$^{\boldsymbol{*}}$ & 14.42$^{\boldsymbol{*}\boldsymbol{\ddagger}}$ & -2.73 & 66.42$^{\boldsymbol{\dagger}\boldsymbol{\ddagger}}$ & 47.10$^{\boldsymbol{\dagger}\boldsymbol{\ddagger}}$ & 15.79$^{\boldsymbol{\dagger}\boldsymbol{\ddagger}}$ & \textbf{+0.45} & & 41.77$^{\boldsymbol{*}\boldsymbol{\ddagger}}$ & 27.79$^{\boldsymbol{*}\boldsymbol{\ddagger}}$ & 11.79$^{\boldsymbol{*}}$ & +4.12 & 61.14$^{\boldsymbol{\dagger}\boldsymbol{\ddagger}}$ & 44.67$^{\boldsymbol{\dagger}\boldsymbol{\ddagger}}$ & 20.99$^{\boldsymbol{\dagger}\boldsymbol{\ddagger}}$ & \textbf{+0.45} \\
& QRACDR-Q$_{MG-MS}$ & 57.11$^{\boldsymbol{*}}$ & 37.80$^{\boldsymbol{*}}$ & 13.59$^{\boldsymbol{*}}$ & -0.21 & 62.77$^{\boldsymbol{\dagger}\boldsymbol{\ddagger}}$ & 44.41$^{\boldsymbol{\dagger}\boldsymbol{\ddagger}}$ & 15.62$^{\boldsymbol{\dagger}\boldsymbol{\ddagger}}$ & \textbf{-5.29} & & 37.79$^{\boldsymbol{*}}$ & 24.94$^{\boldsymbol{*}\boldsymbol{\ddagger}}$ & 11.05$^{\boldsymbol{*}}$ & \textbf{+10.11} & 59.14$^{\boldsymbol{\ddagger}}$ & 43.48$^{\boldsymbol{\ddagger}}$ & 20.26$^{\boldsymbol{\dagger}\boldsymbol{\ddagger}}$ & \textbf{-2.23} \\
\bottomrule
\end{tabular}
}
\end{table*}

\subsubsection{\textbf{Is model merging a comparable training-free alternative to multi-task fine-tuning and early stopping strategies? (RQ2)}}

\begin{figure}[t]
    \centering
    \includegraphics[width=1\linewidth]{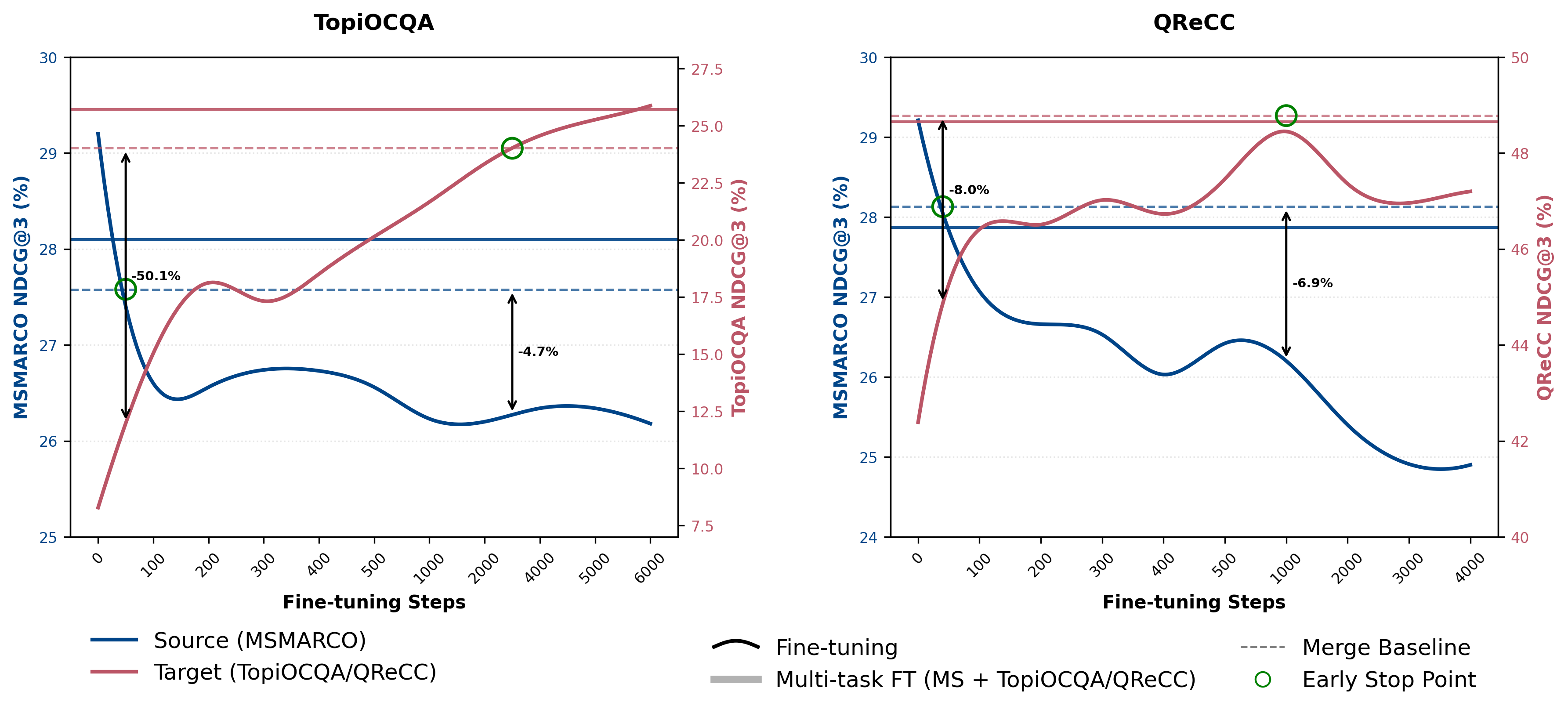}
    \caption{NDCG@3 performance across fine-tuning steps for TopiOCQA (left) and QReCC (right). QRACDR$_{MG-MS}$ (dashed lines) is compared against standard fine-tuning (curves), multi-task fine-tuning (solid lines), and two selected early stopping points (circles). Vertical arrows highlight the performance gap between early stopping models and QRACDR$_{MG-MS}$, demonstrating that model merging effectively preserves source domain (MSMARCO) knowledge while achieving competitive target domain (QReCC/TopiOCQA) performance. Each model  is represented by two curves (blue ad-hoc and red conversational).}
    \label{fig:early-stop}
\end{figure}

We first compare our training-free model merging approach against multi-task fine-tuning (MTL), a standard strategy for mitigating catastrophic forgetting by jointly optimizing over multiple tasks. We adapt the QRACDR training protocol \cite{mo2024aligning}, originally designed for single-dataset learning, to support joint optimization over a balanced mixture of MS MARCO and the conversational target datasets, following the setup proposed by Yang et al.~\cite{yang-etal-2025-learning}. As shown in Fig.~\ref{fig:early-stop}, while MTL alleviates forgetting, our model merging approach matches or exceeds its performance across both conversational and ad-hoc retrieval tasks—without requiring gradient-based optimization or access to the original source training data (-3.7\% NDCG@3 drop for MTL and -5.1\% for QRACDR-T$_{MG-MS}$ in MS MARCO). This makes model merging particularly appealing in scenarios where the source dataset is unavailable or when the size of ad-hoc data renders joint fine-tuning computationally impractical. 

We further compare our method against \emph{early stopping}, a commonly used heuristic for limiting forgetting during sequential fine-tuning.
Fig. \ref{fig:early-stop} reveals a clear conflict during training for both TopiOCQA and QReCC, as the model improves at target conversational dataset, its performance on the original MS MARCO ad-hoc task drops sharply. An intuitive solution is to use an early stopping criteria to limit the number of fine-tuning steps in order to match final performances obtained in one of the tasks. However, our results show that this early stopping criteria is unable to find a satisfactory trade-off. If we stop the training early to keep the ad-hoc search strong, the model fails significantly on the conversational task (e.g., a 50.1\% performance gap on TopiOCQA at step 100). Conversely, if we train long enough to reach high conversational accuracy, at step 3000, the model suffers a 4.7\% loss in its original search effectiveness. Model merging solves this by bypassing the standard training path. By combining the final conversational model with the original search model, we can restore the lost search capabilities without losing the specialized conversational skills. This allows the merged model to achieve a level of balance across both tasks that simple early stopping cannot match.

\subsubsection{\textbf{To what extent does model merging lead to conversational retrievers generalizable across datasets and tasks? (RQ3)}}

In the following experiments, we evaluate the OOD generalizability of the merged models QRACDR-T$_{MG}$ and QRACDR-Q$_{MG}$ on unseen ad-hoc and conversational datasets, using both MS and SL configurations. To quantify generalization abilities, we report the relative performance of our merged models $p(\mathcal{M}_{\theta^{mg}})$ on conversational datasets compared to the performance of the conversational fine-tuned model  $p(QRACDR)$ on each dataset, as
$\delta_i = \frac{p(\mathcal{M}_{\theta^{mg}}) - p(QRACDR)}{p(QRACDR)}$,
and the relative ad-hoc forgetting of the conversational model $p(QRACDR)$ compared to the performance of the base retriever model $p(ANCE)$
$\delta_f = \frac{p(QRACDR) - p(ANCE)}{p(ANCE)} \%$.

\paragraph{\textbf{OOD generalizability across datasets}}

Table \ref{tab:retriever_results_ndcg_only} displays the retrieval performance alongside the relative change ($\delta_f$) for  compared to the baseline \texttt{ANCE} model. The results show that conversational model finetuning results in a performance drop of up to $24.49\%$ on ad-hoc tasks. However, model merging effectively mitigates this forgetting. Notably, \texttt{QRACDR-Q$_{MG-MS}$} achieves a relative improvement of $+4.58\%$ on NQ over the original ad-hoc model, while \texttt{QRACDR-Q$_{MG-SL}$} reduces the severe $19.84\%$ loss on NQ to a $4.42\%$ decrease.

\paragraph{\textbf{OOD generalizability across tasks}}

Here, we evaluate whether merged models can successfully handle both ad hoc and conversational retrieval tasks on respectively standalone and conversational sessions. For this purpose, we use the manually rewritten queries on the CAsT-19 and CAsT-20 datasets. As shown in Table \ref{tab:rewrite_perf}, standard conversational models (QRACDR-T and QRACDR-Q) 

demonstrate strong performance when retrieving from full sessions, but experience a notable performance drop when tested on rewritten queries over the base ANCE model. Specifically, the "Rewrite" performance of QRACDR-Q drops by 15.89\% on CAsT-19. Unlike standard ad-hoc queries from datasets like MS MARCO, which are typically short and keyword-oriented, manually rewritten queries are de-contextualized, standalone questions that explicitly resolve coreferences and ellipsis while maintaining a complex, full-sentence natural language structure. In contrast, our merged models (QRACDR$_{MG}$) bridge this gap by maintaining high effectiveness in both settings. The negative deltas observed in the base conversational models are largely reversed through merging; for instance, QRACDR-T$_{MG}$ eliminates the 8.47\% rewrite loss seen in its conversational counterpart and actually achieves a slight gain over the original ANCE baseline ($\delta_f = +0.45\%$). Furthermore, in the session-based setting for CAsT-20, the merged models show significant improvements over the base conversational versions, with QRACDR-Q$_{MG}$ achieving a +10.11\% increase ($\delta_i$). This suggests that the merging process creates a more robust retriever that is no longer restricted to a single input type, allowing it to function effectively as both a session-based conversational retriever and a high-precision ad-hoc search engine.

\section{Conclusion}
We investigated model merging as a simple approach to restore ad hoc retrieval effectiveness in conversational search models without additional fine-tuning. Compared to multi-task learning and early stopping, model merging achieves a better balance between ad-hoc and conversational performance while exhibiting stronger generalization across datasets and tasks.
These findings suggest that model merging is a promising direction for the design of conversational systems. As future work, we plan to investigate model merging as the pivotal strategy for searching for optimal models based on a scoring function that measures their performance for a specific task  invoked in conversational search, including query clarification,  conversational retrieval, and response generation.


\begin{acks}
The authors acknowledge the ANR – FRANCE (French National Research Agency) for its financial support of the GUIDANCE project
n°ANR-23-IAS1-0003.
This work was granted access to the HPC resources of IDRIS under the allocation AD011016470 made by GENCI.
\end{acks}


\bibliographystyle{ACM-Reference-Format}
\balance
\bibliography{sample-base}


\end{document}